# Highly tunable optical activity in planar achiral terahertz metamaterials

**Ranjan Singh,[1,2,*] Eric Plum,[3] Weili Zhang,[1] and Nikolay I. Zheludev[3]**

[1]*School of Electrical and Computer Engineering, Oklahoma State University, Stillwater, Oklahoma 74078, USA*
[3]*Optoelectronics Research Centre, University of Southampton, SO17 1BJ, UK*
[2]*Center for Integrated Nanotechnologies, Materials Physics and Applications Division,
Los Alamos National Laboratory, Los Alamos, New Mexico 87545, USA*
[*]*ranjan@lanl.gov*

**Abstract:** Using terahertz time domain spectroscopy we demonstrate tunable polarization rotation and circular dichroism in intrinsically non-chiral planar terahertz metamaterials without twofold rotational symmetry. The observed effect is due to extrinsic chirality arising from the mutual orientation of the metamaterial plane and the propagation direction of the incident terahertz wave.

Since the discovery of light polarization rotation in crystals by Arago in 1811 [1], optical activity has been a phenomenon of fundamental importance for the progress of chemistry, physics, biology, and optics. Apart from their ability to rotate the polarization state of light (circular birefringence), optically active materials also exhibit different transmission levels for left and right circular polarizations (circular dichroism) and they can have a negative refractive index for circularly polarized waves [2-4]. Since 1848, when optical activity has been linked to intrinsically 3D-chiral molecules by Luis Pasteur's pioneering work on tartaric acid [5], research in this area focused almost entirely on natural and artificial 3D-chiral structures, i.e. three-dimensional (3D) objects such as a helix, which can be distinguished from their mirror image. However, optical activity does not require intrinsically chiral materials. Instead, optical activity may also arise from extrinsic 3D chirality, when the direction of incidence and an achiral material structure form a chiral experimental arrangement [6-8], see Fig. 1.

The terahertz electromagnetic region is a unique frequency range that is scientifically rich but technologically underdeveloped. It falls between the domain of high frequency radiowaves (microwaves) and the far-infrared with many important applications such as security detection, sensing, biomedical imaging, and in understanding the complex dynamics in solid state physics and processes such as molecular recognition [9,10]. However, the devices for manipulating terahertz waves are considerably limited. Consequently, the development of artificially engineered materials with unusual properties in this frequency region is especially important. Recently, metamaterials research has greatly expanded the accessible range of optical properties at terahertz frequencies [11-28]. However, due to the complexity of 3D-chiral metamaterial geometries, experimental realizations of 3D-chiral terahertz metamaterials still remain challenging. While current realizations of such structures exhibit circular birefringence and circular dichroism [4,29,30], lack of efficient tunability limits their suitability for practical applications.

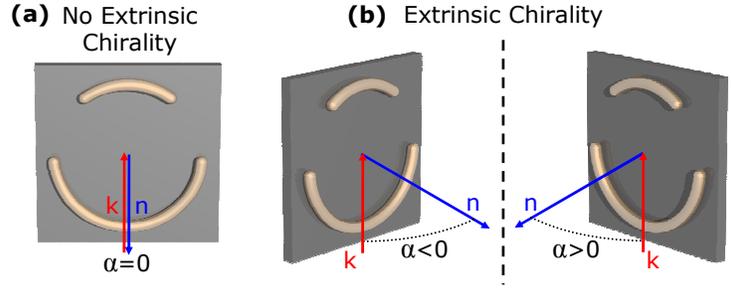

Fig. 1. Extrinsic chirality. (a) At normal incidence, the experimental arrangement formed by the direction of incidence and a planar metamaterial is the same as its mirror image and therefore not 3D-chiral. Here, the angle of incidence α is measured from the metamaterial's surface normal $n$ to the direction of incidence $k$. (b) At oblique incidence, an achiral planar structure and the direction of incidence can form an experimental arrangement that is different from its mirror image and therefore extrinsically 3D-chiral.

Here, we show for the first time that optical activity can occur in non-chiral planar terahertz metamaterials. Exceptionally large circular birefringence and circular dichroism are observed at oblique incidence, when the non-chiral metamaterial forms an extrinsically 3D-chiral experimental arrangement with the incident wave. The effect is inherently tunable, its sign and magnitude can be controlled by the tilt of the metamaterial plane relative to the incident beam. Importantly, this type of tunable optical activity occurs in a large class of simple planar metamaterial designs that are ideally suited for well-established planar manufacturing technologies, opening up an avenue to polarization control devices for terahertz wave applications.

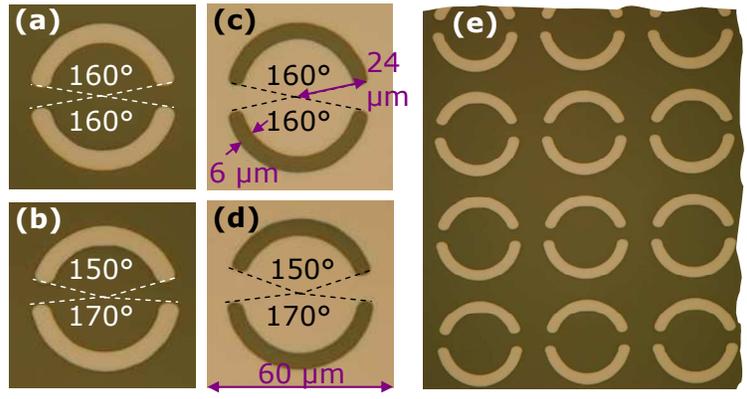

Fig. 2. Fabricated planar metamaterial samples represented by their unit cells: (a) symmetrically and (b) asymmetrically split wire rings and the complementary (c) symmetrically and (d) asymmetrically split ring apertures with structural dimensions. (e) Fragment of the metamaterial array of asymmetrically split wire rings.

Four different planar metamaterials were fabricated by conventional photolithography from a 200 nm aluminum layer on a 640 μm thick silicon substrate (n-type resistivity 12 Ω cm). The metamaterials all have lateral dimensions of 10 x 10 mm$^2$ and a lattice constant of P = 60 μm, which renders the structures non-diffracting at any angle of incidence for frequencies up to 2.5 THz. As illustrated by Figs. 2(a)-2(d), the meta-molecules consist of symmetrically and asymmetrically split wire rings (positive metamaterials) and the complementary symmetrically and asymmetrically split ring apertures in an aluminum film (negative metamaterials), respectively. Detailed dimensions are given in Fig. 2.

We studied these structures using terahertz time-domain spectroscopy THz-TDS [31]. The terahertz beam incident on the sample had a frequency-independent diameter of 3.5 mm and thus illuminated about 3000 unit cells at the center of the metamaterial. Transmission of the structures was studied for angles of incidence in the range from α = -45° to +45°, where the plane of incidence was parallel to the splitting in all cases. Using parallel or crossed linear polarizers placed before and after the sample, we measured all components of the metamaterial's transmission matrix $E_i^{trans} = \tau_{ij} E_j^{inc}$, which relates the incident and transmitted electric fields in terms of linearly polarized components. Amplitude $|\tau_{ij}(\omega)| = |E_{ij}^{sample}(\omega)| / |E^{ref}(\omega)|$ and phase $\arg(\tau_{ij}(\omega)) = \angle\left[ E_{ij}^{sample}(\omega) / E^{ref}(\omega) \right]$ of the transmission matrix elements were calculated from transmission measurements taken on the metamaterial $E_{ij}^{sample}(\omega)$ with a correspondingly oriented blank silicon substrate $E^{ref}(\omega)$ used as reference. The insertion loss of the silicon substrate is about -3 dB. In order to analyze the metamaterial response for circularly polarized waves, we transformed the transmission matrix $\tau_{ij}$ from the linear polarization basis to circular polarization,

$$t = \begin{pmatrix} t_{++} & t_{+-} \\ t_{-+} & t_{--} \end{pmatrix} = \frac{1}{2}\begin{pmatrix} \tau_{xx} + \tau_{yy} + i(\tau_{xy} - \tau_{yx}) & \tau_{xx} - \tau_{yy} - i(\tau_{xy} + \tau_{yx}) \\ \tau_{xx} - \tau_{yy} + i(\tau_{xy} + \tau_{yx}) & \tau_{xx} + \tau_{yy} - i(\tau_{xy} - \tau_{yx}) \end{pmatrix}. \quad (1)$$

In this form the transmission matrix $E_i^{trans} = t_{ij} E_j^{inc}$ directly relates the incident and transmitted terahertz electric fields in terms of right-handed (+) and left-handed (-) circularly polarized components. The square of its elements $T_{ij} = |t_{ij}|^2$ corresponds to transmission and circular polarization conversion in terms of power.

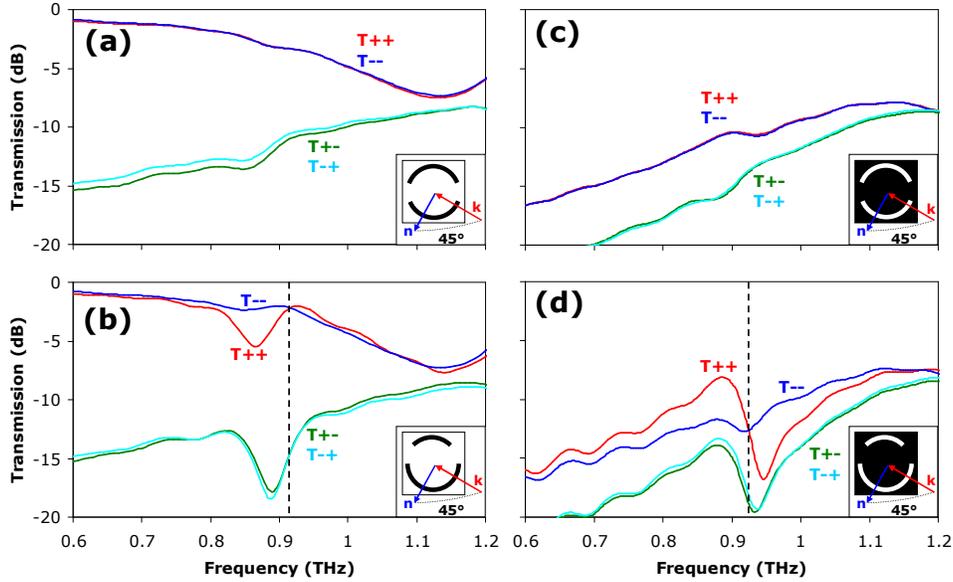

Fig. 3. Measured circular transmission and polarization conversion for terahertz waves incident at α = +45° on metamaterial arrays of (a) symmetrically split wire rings, (b) asymmetrically split wire rings, (c) symmetrically split ring apertures, and (d) asymmetrically split ring apertures. The dashed lines mark the frequency band of pure polarization rotation, where circular dichroism and linear birefringence / dichroism are virtually absent.

Figure 3 shows measurements of direct transmission and circular polarization conversion taken for an angle of incidence of α = 45°. It can be clearly seen, that within experimental accuracy the circular polarization conversion efficiencies are identical for each metamaterial, $T_{-+} = T_{+-}$, indicating the expected presence of linear dichroism/birefringence in the anisotropic structures and absence of the recently-discovered 2D-chiral asymmetric transmission phenomenon [27,32]. More importantly, the direct transmission levels, $T_{++}$ and $T_{--}$, were found to be equal for the symmetric metamaterials and unequal in case of the asymmetric structures, indicating the presence of 3D-chiral circular dichroism, $\Delta = T_{++} - T_{--}$, only for the asymmetric metamaterials. The importance of the asymmetric splitting becomes clear when considering (i) the experimental arrangement consisting of the metamaterial and the direction of incidence and (ii) its mirror image, see Fig. 1(b). For asymmetrically split rings, these mirror-experiments cannot be superimposed: If we rotate one experiment to superimpose the wave vectors $k$ and the surface normals $n$, then the metamaterial orientations will differ by a 180° rotation around $n$ (one structure is up-side-down). Therefore the experimental arrangement is different from its mirror image, extrinsic 3D chirality is present and optical activity is allowed. In case of symmetrically split rings - and indeed for any twofold rotationally symmetric planar metamaterial - the experiment is not chiral, as it can be superimposed with its mirror image and therefore optical activity must be absent.

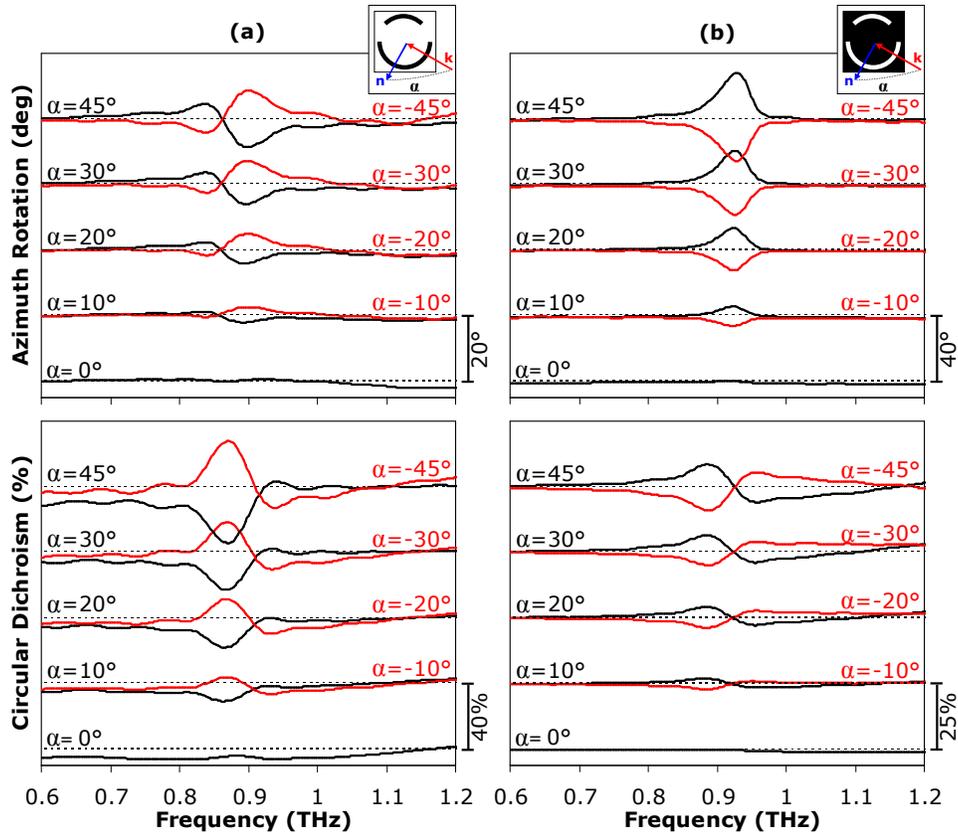

Fig. 4. Optical activity due to extrinsic 3D chirality. Polarization azimuth rotation (circular birefringence) and circular dichroism observed in metamaterial arrays of asymmetrically split (a) wire rings and (b) ring apertures at different angles of incidence α.

The manifestations of extrinsic 3D chirality are further illustrated by Fig. 4, which shows the dependence of polarization rotation, $\Delta\Phi^{3D} = -\frac{1}{2}[\arg(t_{++}) - \arg(t_{--})]$, and circular dichroism Δ on the angle of incidence α for the asymmetric metamaterials. At normal incidence extrinsic chirality is absent and no optical activity can be observed. When the angle of incidence is increased, a band of gradually increasing optical activity appears around 0.9THz, where opposite angles of incidence ±α lead to optical activity of opposite sign. Thus, circular dichroism can be tuned continuously from Δ = -31% to +31%, simply by tilting the array of asymmetrically split wire rings from α = +45° to -45°. Similarly, using the metamaterial array of asymmetrically split ring apertures, polarization rotation can be tuned continuously from $\Delta\Phi^t$ = -28° to +28°. In case of the asymmetric aperture array, it is particularly useful, that the maximum of circular birefringence coincides with absence of circular dichroism and negligible linear dichroism/birefringence [$T_{-+}$, $T_{+-}$, see Fig. 3(d)]. Thus, at the maximum of polarization rotation, the metamaterial exhibits pure circular birefringence and behaves like an ideal tunable polarization rotator. This behavior is opposite to the effect seen in most optically active molecular systems, where characteristically strong resonant polarization rotation of initially linearly polarized radiation is accompanied by substantial circular dichroism resulting in an elliptical polarization state.

Comparing the optical activity for both asymmetric split ring metamaterials, we find that the aperture array shows about 3.5 times larger polarization rotation, see Fig. 4(b). On the other hand, circular dichroism is about 3.5 times larger for the wire structure, see Fig. 4(a). Overall, the spectral and angular dependence of polarization rotation and circular dichroism, each, is similar, but of reversed sign, for the positive and negative metamaterial structures.

Neglecting the presence of the substrate, the positive and negative metamaterials are complementary structures in the sense of Babinet's principle [33-35]. Babinet complementary planar metamaterials exhibit interchanged optical activity in transmission and reflection [36]. Thus it may be expected that both arrays of asymmetrically split rings exhibit substantial polarization rotation and circular dichroism also for reflected waves.

Similarly to optical activity in conventional 3D-chiral molecules, optical activity due to extrinsic 3D chirality has been linked to electric and magnetic responses of the meta-molecules [6]. In a wire split ring, which essentially consists of a pair of electric dipole antennas, in-phase current oscillations correspond to an electric dipole oscillating in the metamaterial plane. On the other hand, the asymmetric splitting also allows the currents in both wires to oscillate in anti-phase, giving rise to a magnetic dipole oscillating perpendicular to the metamaterial plane. Only the magnetic dipole component perpendicular to the propagation direction can contribute to the scattered field and this radiating magnetic component is zero at normal incidence and has opposite signs for opposite angles of incidence. Therefore optical activity can only be observed at oblique incidence onto the asymmetrically split ring arrays and reverses sign for opposite angles of incidence.

In summary, we demonstrate strong and tunable resonant polarization rotation and circular dichroism in achiral planar terahertz metamaterials. The effects are due to extrinsic 3D chirality arising from the mutual orientation of a metamaterial lacking twofold rotational symmetry and the incident terahertz beam. Due to (i) the large magnitude of the observed optical activity, (ii) the huge tunable range of the effect and (iii) the simplicity of suitable planar metamaterials, such structures are ideal functional elements for novel, highly efficient terahertz polarization rotators, circular polarizers, modulators and vibration sensors.

**Acknowledgements**

The authors thank Xinchao Lu for mask design, and Zhen Tian and Jiangfeng Zhou for fruitful discussions. Financial support of the U.S. National Science Foundation and the Engineering and Physical Sciences Research Council, UK are acknowledged.